\documentstyle[11pt,a4]{article}

\newcommand{\be}{\begin{equation}}
\newcommand{\ee}{\end{equation}} 
\newcommand{\bea}{\begin{eqnarray}}
\newcommand{\eea}{\end{eqnarray}} 
\newcommand{\RR}{{\bf {R}}}
\newcommand{\di}{{\rm Diff}S^1 / S^1}
\newcommand{\dis}{{\rm Diff}S^1 / SL^{(n)}(2,\RR)}
\newcommand{\diss}{{\rm Diff}S^1 / SL^{(1)}(2,\RR)}

\catcode`\@=11 
\@addtoreset{equation}{section}
 
\catcode`\@=11

\begin{document}

\begin{titlepage}

\begin{flushright} 
{\tt FTUV/99-23\\ 
     IFIC/99-24 \\ 
     hep-th/9903248}
 \end{flushright}

\bigskip

\begin{center}

{\bf \LARGE Virasoro Orbits, AdS$_3$ Quantum Gravity \\ and Entropy}
\footnote{Work partially supported by the 
{\it Comisi\'on Interministerial de Ciencia y Tecnolog\'{\i}a}\/ 
and {\it DGICYT}.}

\bigskip 

 J.~Navarro-Salas\footnote{\sc jnavarro@lie.uv.es} and
 P.~Navarro\footnote{\sc pnavarro@lie.uv.es}.

\end{center}

\bigskip

\begin{center}
\footnotesize
        Departamento de F\'{\i}sica Te\'orica and 
	IFIC, Centro Mixto Universidad de Valencia-CSIC.
	Facultad de F\'{\i}sica, Universidad de Valencia,	
        Burjassot-46100, Valencia, Spain. 
\end{center}               

\normalsize

\bigskip
\bigskip

\begin{center}
			{\bf Abstract}
\end{center}
We analyse the canonical structure of AdS$_3$ gravity in terms of the
coadjoint orbits of the Virasoro group. There is one subset of orbits,
associated to BTZ black hole solutions, that can be described by a pair
of chiral free fields with a background charge. 
There is also a second subset of orbits, 
associated to point-particle solutions, that 
are described by two pairs of chiral free fields obeying a 
constraint. All these 
orbits admit K\"ahler quantization and generate a Hilbert space which,
despite of having $\Delta_0(\bar{\Delta}_0)=0$,
does not provide the right degeneracy to account for the Bekenstein-Hawking
entropy due to the breakdown of modular invariance.
 Therefore, additional degrees of freedom, reestablishing modular
invariance, are necessarily required
to properly account for the black hole entropy.

\end{titlepage}

\newpage

\section{Introduction}

  Three-dimensional quantum gravity with a negative cosmological constant provides
an interesting example of the general duality relation proposed in 
\cite{Maldacena,GKP,Witten2} between string theory on anti-de Sitter space (AdS) 
times a compact space and a conformal field theory (CFT) on the boundary. It was 
pointed out in \cite{BH} that gravity on AdS$_3$ is a two-dimensional CFT with
a classical central
charge $c_{cl}={3 \over 2}{\ell \over G}$, where G is Newton's constant and
$-{1 \over \ell^2}$ is the cosmological constant. The physical relevance of 2+1
quantum gravity has recently increased \cite{Strominger} since the near-horizon
geometry of black holes arising in string theory can be related to that of the
three-dimensional BTZ black holes \cite{BTZ}. Strominger \cite{Strominger} has 
proposed an unified treatment to account for the Bekenstein-Hawking entropy of all
black holes whose near-horizon geometries are locally AdS$_3$ without using
supersymmetry or string theory. This includes the 
black strings studied in \cite{StVf} as well as the BTZ black holes. The observation
of \cite{Strominger} is based on Cardy's formula \cite{Cardy} for the asymptotic 
density of states of a unitary and modular invariant
 two-dimensional CFT with central charge $c$ and
eigenvalues $\Delta (\bar{\Delta})$ of $L_0 (\bar{L}_0)$
\be
S = 2\pi \sqrt{c \Delta \over 6} +
2\pi \sqrt{c \bar{\Delta} \over 6} \label{Cardy's}
\ee
  As noted in \cite{Strominger} this expression coincides, for 
$\Delta,\bar{\Delta} \gg c$ , with the Bekenstein-Hawking black hole entropy
\be
S = {Area \over 4 G}
\ee
since, for BTZ black holes, one has
\bea
\Delta = {1 \over 2}(\ell M + J) + {\ell \over 16 G} \\
\bar{\Delta} = {1 \over 2}(\ell M - J) + {\ell \over 16 G} 
\eea
  The validity of Cardy's formula (\ref{Cardy's}) requires that the lowest eigenvalues
$(\Delta_0,\bar{\Delta}_0)$ of $L_0$ and $\bar{L}_0$ vanish,
otherwise the asymptotic level density is controlled by the effective central 
charge $c_{eff} = c-24\Delta_0$ \cite{KS}. This hidden assumption for the Cardy's
formula turns out to be very important \cite{Carlip2,Martinec} because it has been
argued \cite{CHvD},using the Chern-Simons formulation of the theory
\cite{AT,Witten},
that the CFT at spatial infinity for AdS$_3$ gravity is Liouville theory.
However, the analysis of \cite{CHvD} is not complete since the zero modes and the
associated holonomies are not considered. Moreover, the lowest eigenvalues
($\Delta_0,\bar{\Delta}_0$) of $L_0$ and $\bar{L_0}$ are not zero
for normalizable states in Liouville theory \cite{Seiberg} 
\be
\Delta_0 = {c-1 \over 24}
\ee
and therefore the central charge in (\ref{Cardy's}) should 
indeed be replaced by $c_{eff}=1$.
This implies that the Liouville theory does not have enough states to account for 
the black hole entropy. Although supersymmetry suggests that the minimum eigenvalue
of $L_0(\bar{L}_0)$ vanishes \cite{Carlip2}, the super-Liouville theory has the
same drawback and fails to give the right degeneracy ($c_{eff} = {3 \over 2}$).
These difficulties were interpreted in \cite{Martinec2} suggesting that gravity
represents a thermodynamical description of the dual CFT with the Liouville field
emerging as a kind of collective coordinate. It has also been argued \cite{GKS,Lee}
that string theory on a AdS$_3$ background could correctly account for the
Bekenstein-Hawking entropy. A recent attempt to attack this problem within gravity
theory has been proposed in \cite{BBB,BBBB} by extending the asymptotic symmetry algebra
with new generators.

  The aim of this paper is to approach this problem from an analysis of the
phase space of the theory in terms of
the coadjoint orbits of the
Virasoro group \cite{Witten3}.

\section{Virasoro orbits and gravity on AdS$_3$}

  To properly define a gravity theory on AdS$_3$ we have to provide boundary
conditions for the fields at infinity. One can assume that the physical metric
field approaches to the AdS$_3$ metric
\be
ds^2 = - ({r^2 \over \ell^2}+1)dt^2+({r^2 \over \ell^2}+1)^{-1}dr^2
+r^2d\theta^2 \, , \label{metric0}
\ee
where $\theta$ and $r$ are the angular and radial coordinates, as follows
\bea
g_{+-}& = & -{r^2\over 2} + {\gamma_{+-}(x^+,x^-)} + {\cal O} ({1 \over r})
 \, , \label{g1} \\  
g_{\pm\pm}& = & {\gamma_{\pm\pm}(x^+,x^-)} + {\cal O}({1 \over r})
 \, , \label{g2} \\
g_{\pm r}& = & {{\gamma_{\pm r}(x^+,x^-)} \over r^3} + {\cal O} ({1 \over r^4})
 \, , \label{g3} \\
g_{rr}& = & {\ell^2 \over r^2} + {{\gamma_{rr}(x^+,x^-)} \over r^4}
 + {\cal O} ({1 \over r^5}) \, , \label{g4}  
\eea
where $x^\pm \equiv {t \over \ell} \pm \theta$. These boundary conditions allow a
well defined action of two copies of the Virasoro group through space-time
diffeomorphisms. The infinitesimal diffeomorphisms $\zeta^a (r, t,\theta)$
preserving the boundary conditions are
\bea
\zeta^+ & = & 2T^+ + {\ell^2\over r^2} \partial_-^2 T^- + {\cal O} ({1\over r^4})
 \, , \label{d1} \\ 
\zeta^- & = & 2T^- +{\ell^2 \over r^2} \partial_+^2 T^+ + {\cal O} ({1 \over r^4})
 \, , \label{d2} \\
\zeta^r & = & -r(\partial_+ T^+ + \partial_- T^-) + {\cal O} ({1\over r}) \, ,
\label{d3}
\eea
where the functions $T^\pm$ depend on $x^{\pm}$ ($T^\pm(r,t,\theta)
 = T^\pm(x^{\pm}$)).
   By using diffeomorphisms which can be regarded as "gauge transformations"
($T^\pm=0$) one can bring a general metric satisfying the equations of motion
\be
R_{\mu\nu} - {1 \over 2}g_{\mu\nu}R = -{1 \over \ell^2}g_{\mu\nu}
\ee
into the form \cite{ads3}
\be
ds^2 = {\ell^2 \over r^2} dr^2 - r^2 dx^+dx^- + \gamma_{++}(dx^+)^2 +
\gamma_{--}(dx^-)^2 + 
{\cal O} ({1 \over r}) \, . \label{metric}
\ee
where $\gamma_{\pm}(x^{\pm})$ are chiral functions. If either $\gamma_{++}=0$ or
$\gamma_{--}=0$, the omitted corrections vanish. Recently \cite{max}, it has been
obtained an exact general solution, which can be rewritten as
\be
ds^2 = {\ell^2 \over r^2} dr^2 - (rdx^- -{\gamma_{++} \over r}dx^+)
(rdx^+ -{\gamma_{--} \over r}dx^-).
\ee 
  A special class of solutions verifying the boundary conditions (\ref{g1}-\ref{g4})
are the BTZ black holes \cite{BTZ}. They correspond to constant functions for
$\gamma_{++},\gamma_{--}$:
\be
\gamma_{\pm\pm} = 2G\ell(M\ell \pm J) \label{blackhole}
\ee
where $M$ is the black hole mass and $J$ the angular momentum.

  The physical excitations can be naturally defined by the action of the
"would-be gauge" diffeomorphisms on the topologically inequivalent geometries to
AdS$_3$ (\ref{metric0}). Obviously, the geometries obtained by a discrete 
identification of AdS$_3$ cannot be related by diffeomorphisms and therefore
the physical phase space ${\cal M}$ of the theory is the collection of the
diffeomorphism orbits through the topologically different solutions. 
We shall now provide a detailed description of this phase space.
To this end we have
to know the action of the diffeomorphisms on the solutions (\ref{metric}).

  The action of the infinitesimal diffeomorphisms (\ref{d1}-\ref{d3}) 
on the functions
$\gamma_{\pm\pm}$ is
\be
\delta_{T^\pm}\gamma_{\pm\pm} = 2(T^\pm \partial_{\pm} \gamma_{\pm\pm} 
+ 2\gamma_{\pm\pm} \partial_{\pm} T^\pm) - \ell^2 \partial_{\pm}^3 T^{\pm} \, .
\label{delta}
\ee
and the Noether charges $J[\xi]$ associated with them are  
 \be
J[\xi] = {1 \over 16\ell G}\int d\phi \lbrace T^+(4\gamma_{++} + \ell^2) +
T^-(4\gamma_{--} + \ell^2) \rbrace \, ,
\ee
  These expressions allow to relate $\gamma_{\pm\pm}$ to the stress tensor
$\Theta_{\pm\pm}$ of a conformal field theory on the sphere:
$\Theta_{\pm\pm}={1 \over 4\ell G}\gamma_{\pm\pm}$.
With this identification the transformation law 
(\ref{delta}) for  $\Theta_{\pm\pm}$ is

\be
\delta_{T^\pm}\Theta_{\pm\pm} = 2(T^\pm \partial_{\pm} \Theta_{\pm\pm} 
+ 2\Theta_{\pm\pm} \partial_{\pm} T^\pm) - {c \over 12}
 \partial_{\pm}^3 T^{\pm} \, , 
\ee
where the classical central charge can be worked out immediately 
\be
c_{cl} = {3 \over 2}{\ell \over G}
\ee
  The Fourier components $L_n(\bar{L}_n)$ of $\Theta_{++}(\Theta_{--})$
close down the Virasoro algebra in the Ramond form
\bea
i \left\{ L_n,L_m\right\} & = & (n-m)L_{n+m} + {c \over 12}n^3\delta_{n,-m} \, ,
\\
i \left\{ \bar{L}_n,\bar{L}_m\right\} & = & (n-m)\bar{L}_{n+m} +
 {c \over 12}n^3\delta_{n,-m} \, , \\
\left\{ L_n,\bar{L}_m\right\} & = & 0 \, ,
\eea

  The integrated form of (\ref{delta}) is
\be
\gamma_{\pm\pm}\longrightarrow (\partial_{\pm}F_{\pm})^2 \gamma_{\pm\pm}
-{\ell^2 \over 2}(\frac{\partial_{\pm}^3 F_{\pm}}{\partial_{\pm}F_{\pm}}
-{3 \over 2}(\frac{\partial_{\pm}^2 F_{\pm}}{\partial_{\pm}F_{\pm}})^2)
\label{trnsf}
\ee
where $F_{\pm}(x^{\pm})$ are diffeomorphisms of the sphere $S^1$
parametrized by $e^{ix^{\pm}}$. The expression (\ref{trnsf})
turns out to be equivalent to the coadjoint action $Ad^{\ast}(F_{\pm})$ of
the Virasoro group \cite{Witten3}
 and therefore the contribution of (\ref{trnsf}) to the phase
space can be identified with some coadjoint orbit of the 
Virasoro group  ${\rm Diff}S^1 / H$, where $H$ is the stationary subgroup of the
orbit. The most interesting orbits \cite{Witten3} emerge when $\gamma_{\pm\pm}$ are
constant, and this is the case in our theory (see (\ref{blackhole})).
For a generic constant value
$\gamma_{\pm\pm} = 8\pi\ell G b_0^{\pm}$ 
the subgroup $H$ is the rotation group $S^1$ and the
coadjoint orbit is then $\di$. However, for special values of  $b_0^{\pm}$
\be
b_0^{\pm} = -n^2{c_{cl} \over 48\pi}
\ee
$n = 1,2,3,\dots \, ,$ the stationary subgroup becomes larger
\be
H = \dis
\ee
where $SL^{(n)}(2,\RR)$ is generated by $L_0, L_n$ y $L_{-n}$.
Since the minimum value of $b_0^{\pm}$ is given by anti-de Sitter space 
($b_0^{\pm}=-{c_{cl} \over 48\pi}$) the relevant orbits
of our problem are the following
\bea
b_0^{\pm}& = & -{c_{cl} \over 48\pi}  \qquad \diss\oplus\diss \\
b_0^{\pm}& > & -{c_{cl} \over 48\pi}  \qquad \di\oplus\di
\eea
The sector $b_0^{\pm}<0$ corresponds to classical point-particle solutions
\cite{DeserJ}.

  These orbits naturally inherit the symplectic two-form \cite{Witten3,AS,ANN}
\be
\omega = \omega_{+} + \omega_{-}
\ee
where
\be
\omega_{\pm} = {c_{cl} \over 48 \pi}\delta\int_0^{2\pi}dx^{\pm}
\frac{\delta\partial_{\pm}^2F_{\pm}}{\partial_{\pm}F_{\pm}}
+ b_0^{\pm}\delta\int_0^{2\pi}dx^{\pm}
\partial_{\pm}F_{\pm}\delta F_{\pm}
\label{omega1}
\ee
  Note that, for convenience, we are still using the group variables to
parametrize the orbits.

  The Fourier expansion of $F_{\pm}$
\be
F_{\pm}(x^{\pm}) = x^{\pm} + {1 \over 2\pi}\sum_{k\ne 0}s_{k}^{\pm}
e^{-ikx^{\pm}}
\ee
implies that
\be
\omega_{\pm} = -{i \over 24}c(k^3-n^2 k)\delta s_{-k}^{\pm}
\wedge\delta s_{k}^{\pm} + {\cal O}(s)
\ee
So, to lowest order in a $1/c_{cl}$ expansion the Poisson brackets of $s_k(k\ne 0)$
are
\be
\left\{s_{k}^{\pm},s_{r}^{\pm}\right\} = i{24 \over c}(k^3-n^2 k)^{-1}
\delta_{k,-r} + {\cal O}({1 \over c_{cl}^2})
\ee
which are similar to the Poisson brackets of free bosons. However, it is well-known
that a symplectic structure can always be written, at least locally, in the
standard form $\omega = \sum_{i}\delta p_{i}\wedge\delta q_{i}$. In our case,
the natural ansatz for the Darboux fields $\phi_{\pm}$ is
\be
\phi_{\pm} = \sqrt{{c_{cl} \over 3}}({1 \over 2}\ln\partial_{\pm}F_{\pm}
+ \alpha_{\pm}F_{\pm}) \, ,
\ee
where $\alpha_{\pm}$ are arbitrary real parameters.
This gives a symplectic form
\bea 
\omega_{\pm} & = & {1 \over 4\pi}
\delta\int_0^{2\pi}dx^{\pm}\partial_{\pm}\phi_{\pm}\delta\phi_{\pm} \\
& = & {c_{cl} \over 48\pi}\delta\int_0^{2\pi}dx^{\pm}
\frac{\delta\partial_{\pm}^2F_{\pm}}{\partial_{\pm}F_{\pm}}
+ {\ell\alpha_{\pm}^2 \over 8\pi G}\delta\int_0^{2\pi}dx^{\pm}
\partial_{\pm}F_{\pm}\delta F_{\pm} \, .
\eea

For $b_0^{\pm} \ge 0$,
in order to recover (\ref{omega1}), we must choose
\be
\alpha_{\pm} = \sqrt{{8\pi G \over \ell}b_0^{\pm}}
\ee
Then, the stress tensor takes the form of improved chiral free fields
\be
\Theta_{\pm\pm} = {1 \over 2}((\partial_{\pm}\phi_{\pm})^2
-\sqrt{\ell \over 2G}\partial_{\pm}^2\phi_{\pm})
\ee
So, the subset of orbits $b_0^{\pm} \ge 0$ can be described in terms of
a pair of chiral fields $\phi_{\pm}$ whose zero-modes are related to the
$b_0^{\pm}$ parameters through
\be
\phi_{\pm}(x^{\pm}\pm 2\pi) = \phi_{\pm}(x^{\pm}) \pm 2\pi\sqrt{4\pi b_0^{\pm}}
\ee
  We should note that, in case that $b_0^{+}=b_0^{-}$ (i.e., $J=0$), the left and
right moving sectors can be summed up to produce a scalar free field
$\phi = \phi_{+} + \phi_{-}$ which, in turn, can be mapped via a B\"acklund 
transformation into a Liouville field \cite{BCT,HJ}
\be
\phi_{L} = \sqrt{c_{cl} \over 3}({1 \over 2}\ln \frac{\partial_{+}A_{+}
\partial_{-}A_{-}}{(1+{\lambda^2 \over 2}A_{+}A_{-})^2}) \, , \label{Liouville}
\ee
where $A_{\pm}=F_{\pm}$ if $\alpha_{+}=\alpha_{-}=0$ and
$A_{\pm}={1 \over 2\alpha_{\pm}}e^{2\alpha_{\pm}F_{\pm}}$ 
if $\alpha_{+}=\alpha_{-}\neq 0$, and $\lambda^2$ is an arbitrary constant. 
In this situation
we have\footnote{The B\"acklund transformation defines a proper canonical 
transformation
only if the monodromy of the chiral functions $A_{\pm}$ of the Liouville field
is hyperbolic ($\alpha_{+}=\alpha_{-}\neq 0$) or parabolic ($\alpha_{+}=\alpha_{-}
= 0$).}

\be
\omega = \omega_{+} + \omega_{-} = {1 \over 2\pi}\int_{0}^{2\pi}dx
\delta\dot{\phi_{L}}\delta\phi_{L} \label{sym}
\ee
and
\be
\Theta_{\pm\pm} = {1 \over 2}[(\partial_{\pm}\phi_{L})^2-\sqrt{\ell \over 2G}
\partial_{\pm}^2\phi_{L}] \label{sttensor}
\ee
  This way we recover the results of \cite{CHvD} obtained from the Chern-Simons gauge
theory by implementing a Hamiltonian reduction of two chiral WZW models.
However, the analysis of \cite{CHvD} does not consider the zero modes and,
therefore, cannot see the details of all the Virasoro orbits.
In fact, only for $J=0$ one can construct a Liouville field from the chiral free
fields $\phi_{\pm}$.

The situation for $b_0^{\pm}<0$ is more involved.
It would be necessary to choose $\alpha_{\pm}$ 
imaginary, so that these orbits cannot be described in terms of a 
pair of real chiral
fields $\phi_{\pm}$ only. In general we have:

\be
\Theta_{\pm\pm} = {1 \over 2}\{(\partial_{\pm}\phi_{\pm})^2
-\sqrt{\ell \over 2G}\partial_{\pm}^2\phi_{\pm}\} +
(2\pi(b_0^{\pm})^2-{\ell\alpha_{\pm}^2 \over 4G})(\partial_{\pm}F_{\pm})^2
\ee
  Nevertheless, we can describe the subset of orbits $-{c_{cl} \over 48\pi} \le
b_0^{\pm} <0$
by two pairs of chiral fields $\varphi_{\pm},\eta_{\pm}$

\bea
\sqrt{3 \over c_{cl}}\varphi_{\pm} & = & {1 \over 2} \ln{\partial_{\pm}F_{\pm}} \\
\eta_{\pm} & = & F_{\pm}
\eea
obeying the constraint
\be
e^{2\sqrt{3 \over c}\varphi_{\pm}} = \partial_{\pm}\eta_{\pm} \label{constraint}
\ee
The symplectic form is
\be
\omega_{\pm} = {1 \over 4\pi}\delta\int_0^{2\pi}dx^{\pm}\partial_{\pm}
\varphi_{\pm}\delta\varphi_{\pm} +
b_0^{\pm}\delta\int_0^{2\pi}dx^{\pm}\partial_{\pm}
\eta_{\pm}\delta\eta_{\pm}
\ee
and the stress tensor becomes
\be
\Theta_{\pm\pm} = {1 \over 2}[(\partial_{\pm}\varphi_{\pm})^2
-\sqrt{\ell \over 2G}\partial_{\pm}^2\varphi_{\pm}]
 + 2\pi b_0^{\pm}(\partial_{\pm}\eta_{\pm})^2 \, . \label{stress}
\ee
  In conclusion, the canonical structure of the orbits with $b_0^{\pm}<0$ is
captured by two pairs of chiral free fields with indefinite signature, one pair
with a background charge and the other without improvement, verifying a constraint.
We must note that, in contrast with the subset of orbits $b_0^{\pm}\ge 0$,
the presence of the parameters $b_0^{\pm}$ in the stress tensor (\ref{stress})
makes difficult to derive it from a boundary action. This fact will be reflected
in the absence of modular invariance in the contribution of these orbits to
the Hilbert space.

One can equivalently describe the orbits $-{c_{cl} \over 48\pi} \le
b_0^{\pm} <0$ in terms of a complex
chiral scalar field
\be
\phi_{\pm}=\sqrt{c_{cl} \over 3}({1 \over 2}\ln\partial_{\pm}F_{\pm}+
i\vert\alpha_{\pm}\vert F_{\pm})
\ee
which can be rewritten as
\be
\phi_{\pm}=\sqrt{c_{cl} \over 3}{1 \over 2}\ln \partial_{\pm}A_{\pm}
\ee
where
\be
A_{\pm}={1 \over 2i\vert\alpha_{\pm}\vert}e^{2i\vert\alpha_{\pm}\vert F_{\pm}}
\ee
  The fields $A_{\pm}$ have elliptic monodromy
\be
A_{\pm}(x^{\pm} \pm 2\pi) = e^{\pm\theta_{\pm}i}A_{\pm}(x^{\pm})
\ee
where $\theta_{\pm}=4\pi\vert\alpha_{\pm}\vert\in ]0,2\pi]$.
Note that anti-de Sitter space ($b_0^+ = b_0^- = -{c_{cl} \over 48\pi}$)
corresponds to the trivial monodromy $\theta_{\pm}=2\pi$.
For spinless solutions ($J=0$), i.e. $\theta_+ = \theta_-$, we can again
join the two chiral sectors to give a real Liouville field, like in
(\ref{Liouville}),
with symplectic form and stress tensor given by (\ref{sym}) and (\ref{sttensor}).

\section{Quantization}
  
\setcounter{footnote}{1}
  We have seen that the subset of the
Virasoro orbits with $b_0^{\pm}\ge 0$ are described
by two chiral free fields with a classical background charge $Q_{cl}=
\sqrt{c_{cl} \over 3}$ and this allows a straightforward quantization.
At the quantum level
\be
Q = \sqrt{c_{cl} \over 3} + 2\sqrt{3 \over c_{cl}}
\ee
and the central charge is
\be
c = 1 + 3Q^2 \label{cQ}
\ee
  Due to the presence of the background charge, the state-operator correspondence
for momentum eigenstates have the following form
\be
e^{ip_{\pm}\varphi_{\pm}(0)+{Q \over 2}\varphi_{\pm}(0)}\mid 0 \rangle
= \mid p_{\pm} \rangle
\ee
and it follows that
\bea
L_{0}\mid p_{\pm}\rangle = {Q^2 \over 8}+{p_{+}^2 \over 2}\mid p_{+}\rangle \\  
\bar{L}_0\mid p_{-}\rangle = {Q^2 \over 8}+{p_{-}^2 \over 2}\mid p_{-}\rangle
\eea
  Therefore the contribution of this sector to the Hilbert space is
\be
\bigoplus_{\Delta,\bar{\Delta}\ge{c-1 \over 24}}
{\cal H}_{\Delta}\otimes{\cal H}_{\bar{\Delta}} \label{1stsector}
\ee
where ${\cal H}_{\Delta}$(${\cal H}_{\bar{\Delta}}$) are Virasoro
representations with lowest $L_0(\bar{L}_0)$\footnote{These
 are now the usual Neveu-Schwarz $L_0$($\bar{L}_0$) 
operators.}
eigenvalue $\Delta(\bar{\Delta})$ and central charge (\ref{cQ}).

  In contrast, for $-{c_{cl} \over 48\pi} \le b_0^{\pm}<0$
the canonical structure of the
orbits is more involved. We need two pairs of chiral fields ($\varphi_{\pm}$,
$\eta_{\pm}$) obeying the constraint (\ref{constraint}).
The quantization can be carried out
consistently imposing the following condition on the physical states $\mid\psi\rangle$ 
\be
\langle\psi\mid :e^{2\sqrt{3 \over c}\varphi_{\pm}}: 
- :\partial_{\pm}\eta_{\pm}:\mid\psi\rangle = 0 \label{constraint2} 
\ee
  It is important to note that the two terms of the quantum constraint
are primary fields with the same conformal weight.
 The first term is a chiral vertex operator
$:e^{2\alpha\phi}:$ with a conformal dimension 
$\Delta(\bar{\Delta})=-{1 \over 2}\alpha^2
+{1 \over 2}\alpha Q$, due to the presence of the background charge 
$Q = \sqrt{c_{cl} \over 3}+2\sqrt{3 \over c_{cl}}$.
A simple calculation gives $\Delta(\bar{\Delta})=1$ and coincides with the dimension of the
second term since $\eta_{\pm}$ are free fields without 
improvements. Moreover, the quantum Virasoro algebras are generated by
the operator version
of (\ref{stress}), which weakly commute with the constraint
giving rise, in the light of the AdS/CFT duality, to the central charge (\ref{cQ}).
Therefore, the quantum central charge of 2+1 gravity coincides with
 that of Liouville theory
with classical central charge $c_{cl}={3 \over 2}{\ell \over G}$. However,
the full Hilbert space is not isomorphic to that of
quantum Liouville theory. The
sector coming from the classical point-particle solutions add to the Hilbert
space the following Virasoro representations
\be
\bigoplus_{{c-1 \over 24}>\Delta,\bar{\Delta}\ge 0}
{\cal H}_{\Delta}\otimes{\cal H}_{\bar{\Delta}}
\ee

  From the geometrical point of view these results are consistent whit the
fact that all orbits with $b_0^{\pm}\ge -{c_{cl} \over 48\pi}$ can be
quantized because they posses a K\"ahler structure \cite{Witten3}.
It is interesting to
remark that all the orbits $\di$ are K\"ahler manifolds \cite{B-R}, but only
$\dis$ with $n=1$ admit a K\"ahler structure \cite{Witten3}.
In other words, only $AdS_3$
($b_0^{\pm}=-{c_{cl} \over 48\pi}$) generate a quantizable orbit with a
$SL(2,\RR)$ symmetry.

\section{Conclusions and final comments}

  We have shown that the phase space of $AdS_3$ gravity, with the Brown-Henneaux
boundary conditions, can be described in terms of the coadjoint orbits of
the Virasoro group and splits into two sectors. 
The sector associated to classical black hole solutions is described by a pair
of chiral free fields with a background charge giving rise to a quantum central charge
equal to that of a Liouville theory. 
However, only when $J=0$ the two chiral free fields can be summed up to produce a
scalar field which can also be mapped, through a canonical transformation, into
a Liouville field with hyperbolic and parabolic solutions.
The second sector requires two pairs of chiral free
fields, one of then with a background charge, obeying a special constraint.
Moreover, this sector is canonically equivalent, for $J=0$, to a classical Liouville
field with elliptic monodromy. Nevertheless it is important to point out that,
although the classical solutions of three-dimensional gravity with $J=0$ can be 
associated to classical solutions of Liouville theory, 
we have seen that the primary description
of the gravity theory appears in terms of chiral free fields and therefore the
correspondence with Liouville theory is not valid at the quantum level,
as it has been suggested in  \cite{Martinec2}. Only
remains true in the first sector, which can be related to normalizable solutions
of quantum Liouville theory \cite{Seiberg}.  
The set of Virasoro representations emerging in this sector (\ref{1stsector})     
is modular invariant, but this is no longer true for the second sector and Cardy's
formula does not apply. This can be checked immediately because, by direct counting,
the asymptotic density of states is the same as in the first sector and it is
controlled by $c_{eff}=1$. Therefore, a Hilbert space of a CFT which could
explain the Bekenstein-Hawking entropy requires an enlarged Hilbert space 
\be
\bigoplus_{\Delta,\bar{\Delta}\ge 0}N_{\Delta,\bar{\Delta}}
{\cal H}_{\Delta}\otimes{\cal H}_{\bar{\Delta}} \, ,
\ee
where the positive integer coefficients $N_{\Delta,\bar{\Delta}}$, which
stand for the multiplicities of the corresponding Virasoro representations, are
such that ensures modular invariance. The gravity theory
with the standard Brown-Henneaux boundary conditions is able to see the
different Virasoro representations entering the Hilbert space, but not the 
corresponding multiplicities. This work is left to an additional microscopic
structure like string theory \cite{GKS,Lee}
or the twisted states recently introduced in \cite{BBBB}.

The recent paper \cite{NUY} also analyzes three-dimensional gravity using
the coadjoint orbits of the Virasoro group.

\section*{Acknowledgements}
We would like to thank M. Ba\~nados and A. Mikovic for useful discussions. 
P. Navarro acknowledges the Ministerio de Educaci\'on y Cultura for a FPU fellowship.

\end{document}